# Generation of rotational ground state HD$^+$ ions in an ion trap using a resonance-enhanced threshold photoionization process

Yong Zhang[1,2], Qian-Yu Zhang[1,2], Wen-Li Bai[1,2], Zhi-Yuan Ao[1,2], Wen-Cui Peng[1],
Sheng-Guo He[1,*], Xin Tong[1,3,**]

[1] *State Key Laboratory of Magnetic Resonance and Atomic and Molecular Physics, Innovation Academy for Precision Measurement Science and Technology, Chinese Academy of Sciences, Wuhan 430071, People's Republic of China*

[2] *University of Chinese Academy of Sciences, Beijing 100049, People's Republic of China*

[3] *Wuhan Institute of Quantum Technology, Wuhan 430206, People's Republic of China*

We report a method for producing ultracold HD$^+$ molecular ions populated in a rotational ground state in an ion trap based on [2+1'] resonance-enhanced threshold photoionization (RETPI) and sympathetic cooling with the laser-cooled Be$^+$ ions. The effect of electric field of the ion trap on the RETPI process of neutral HD molecules and the blackbody radiation (BBR) on the population evolution of rotational states of the generated polar HD$^+$ ions have been studied. The initial rotational ground state population of HD$^+$ ions is 0.93(12). After the cumulation time of 5 s, the rotational ground state population is reduced to 0.77(8) due to the BBR coupling. This method of generating ultracold state-selected HD$^+$ ions is beneficial for the studies in precision rovibrational spectroscopy, state-controlled cold chemical reaction, and quantum logic spectroscopy.

## I. INTRODUCTION

The hydrogen molecular ion (HD$^+$) is the simplest heteronuclear molecular ion, and its rovibrational transition frequencies can be accurately obtained by both theoretical calculation and experimental measurement [1-4]. Therefore, HD$^+$ ion is a benchmark system for molecular theory and a suitable probe for fundamental physical models [5-10]. The key to tapping the full potential of molecular ions is the ability to accurately control the external and internal degrees of freedom (DOF) of HD$^+$ ions. The external DOF of HD$^+$ ions can be sympathetically cooled to the order of millikelvin by the co-trapped laser-cooled atomic ions using Coulomb interaction [11-14]. However, due to the uncoupling of the internal and external DOF, the populations of the internal states still maintain the thermal equilibrium distribution of ambient temperature. For HD$^+$ ions at room temperature, almost all the ions are populated in the $v^+ = 0$ vibrational state of the 1s$\sigma$ electronic ground state, but only 10% are populated in the $v^+ = 0, J^+ = 0$ rovibrational ground state. The remaining ions are distributed mostly among the $v^+ = 0, J^+ = 1 - 5$ rotational excited states [15]. Therefore, the internal state preparation of HD$^+$ molecular ions is needed to improve the signal-to-noise ratio in precise spectral measurements [16-17].

In recent years, many schemes for internal state preparation of molecular ions have been realized, including cryogenic buffer-gas cooling [18-22], collision with cold atom cloud [23-30], optical pumping [31-36], and resonance-enhanced threshold photoionization (RETPI) [37-38]. By interacting with the helium buffer gas, a single molecular ion MgH$^+$ is cooled to an internal temperature of 7.5 kelvin [18-19]. However, the cooling is at the expense of the translational temperature, and the MgH$^+$ ion is heated to a few hundred millikelvin. Rotational or vibrational quenching in collisions between molecular ions and cold or ultracold neutral atoms have been studied theoretically [23-26]. So far, only some evidence of vibrational quenching, rather than rotational quenching, of molecular ions has been observed in hybrid atom-ion traps [27-30]. The optical pumping of selected electronic or rovibrational transitions can accumulate the population of the ground state. However, the former is only applied to vibrational-electronic decoupled molecular ions, such as SiO$^+$ and AlH$^+$ [31-32], and the latter requires a complex laser system to excite corresponding rovibrational transitions [15, 33]. RETPI, in principle, can be applied to a wide range of molecules, including apolar and polar molecules. This method has been demonstrated to generate apolar molecular ions N$_2^+$ with a population of 93% in the selected rotational state [37]. However, it has not been demonstrated to generate polar molecular ions experimentally. Even if polar molecular ions are generated, the lifetime of state-selected ions also depends on the molecular dipole moment, rotational constants and blackbody radiation (BBR) intensity [15].

In this paper, we report on preparing rotational ground state HD$^+$ ions in an ion trap by [2+1'] RETPI via a selected $EF\ ^1\Sigma_g^+, v' = 0, J' = 0$ intermediate state of neutral HD molecules. The rotational state selection in HD$^+$ ions is achieved by setting the ionization laser frequency slightly above the lowest rotational ionization thresholds accessible from the intermediate state [Fig. 1(a)]. Due to the indistinguishability of HD$^+$ ions produced by direct photoionization and field ionization in the ion trap, the production of HD$^+$ ions is performed first in a time-of-flight mass spectrometer (TOFMS), while both direct current (dc) and radio frequency (rf) electric field are applied to the electrode plates of the TOFMS to mimic the electric trapping field of the ion trap. The frequency of the threshold ionization laser is determined by considering the influence of the dc and rf electric fields of the ion trap on the ionization process of neutral HD molecules. Then, the rotational ground state HD$^+$ ions are generated in the ion trap and sympathetically cooled by co-trapped laser-cooled Be$^+$ ions. The rotational ground state population of HD$^+$ ions is derived by the results of the state-selected [1+1'] resonance-enhanced multiphoton dissociation (REMPD) [Fig. 1(c)] combined with the rate equations modeling the evolution of the rotational state populations due to BBR coupling.

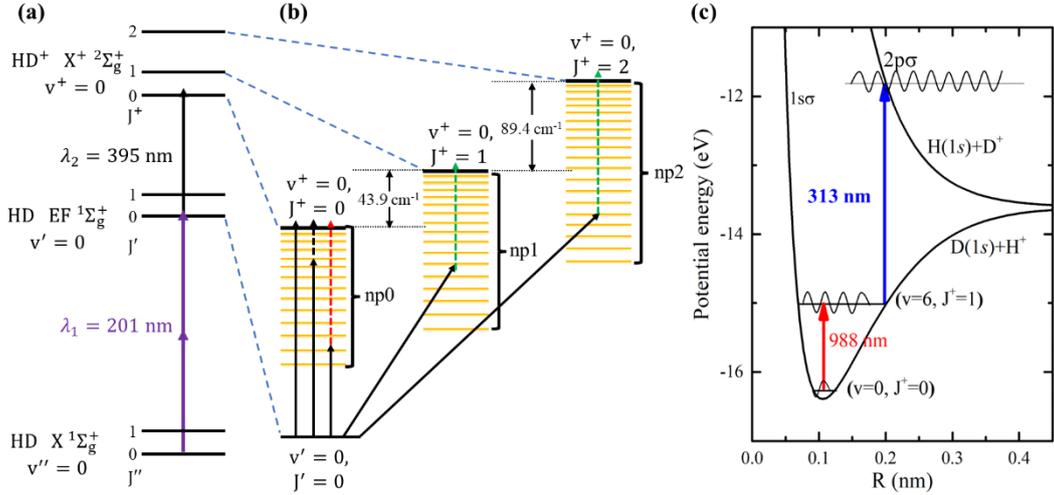

**FIG. 1**. A scheme for production and detection of state-selected HD$^+$ ions. (a) [2+1'] RETPI scheme for producing state-selected HD$^+$ ions; (b) The detailed partial level scheme shows the ionization processes with the existence of electric field, including photoionization and electric field ionization. np0 (np1, np2) is a series of Rydberg states converging to the $J^+ = 0$ ($J^+ = 1, 2$) rotational state of HD$^+$ ions. The solid arrows represent the transitions excited by the ionization laser ($\lambda_2$). The dashed arrows indicate electric field ionization processes of creating prompt ions (in black), mass-analyzed threshold ionization (MATI) ions (in red), and state-unselected ions (in green). (c) State-selected detection of HD$^+$ ions by [1+1'] photon dissociation.

## II. EXPERIMENTAL SETUP

The experimental setup is shown in Fig. 2(a). The apparatus, including a pulsed HD molecular beam apparatus, a time-of-flight mass spectrometer (TOFMS) and an ion trap, have been described in detail in our previous works [39-42]. HD$^+$ ions are generated in the ion trap by an excitation laser ($\lambda_1 \approx 201$ nm, 0.08 mJ/pulse) and an ionization laser ($\lambda_2 \approx 395$ nm, 1.5 mJ/pulse). The 201 nm and 395 nm lasers with the spectral linewidths of 0.4 cm$^{-1}$ and 0.2 cm$^{-1}$, respectively, are the frequency-tripled and the frequency-doubled output from two dye lasers pumped by a commercial Nd:YAG laser with a repetition rate of 10 Hz and pulsed width of $\tau_p = 8$ ns. The created HD$^+$ ions are sympathetically cooled by the co-trapped Be$^+$ ions which are Doppler cooled by a 313 nm laser. For probing the rotational ground state population, HD$^+$ ions are excited to the $v^+ = 6, J^+ = 1$ rovibrationally excited state by a 988 nm laser ($I_{988} \approx 40$ mW/mm$^2$), and the rovibrational transition frequency is precisely calculated [43-45]. The vibrationally excited ions are further dissociated with the rate of 40 s$^{-1}$ by the 313 nm laser ($I_{313} \approx 30$ mW/mm$^2$) through excitation to the 2p$\sigma$ electric excited state without affecting the ions populated in the $v^+ = 0$ state [46].

The structure of the segmented linear ion trap used in our experiment can be inferred from Ref. [42]. For trapping HD$^+$ and Be$^+$ ions simultaneously, a rf amplitude of $V_{rf, p-p} = 600$ V at $2\pi \times 15.1$ MHz and a dc voltage of $V_{end} = 1.5$ V are applied. The maximal electric field strength distribution near the center of the ion trap, where photoionization takes place, is simulated by electromagnetic analysis [Fig. 2(b)]. Within a radius of 0.5 mm from the center of the ion trap, the electric field strength is less than 35 V/cm in the radial direction and kept constant within 3 V/cm in the axial direction. In addition, the Coulomb field induced by the surrounding ions, which is less than 0.1 V/cm, is negligible in this paper.

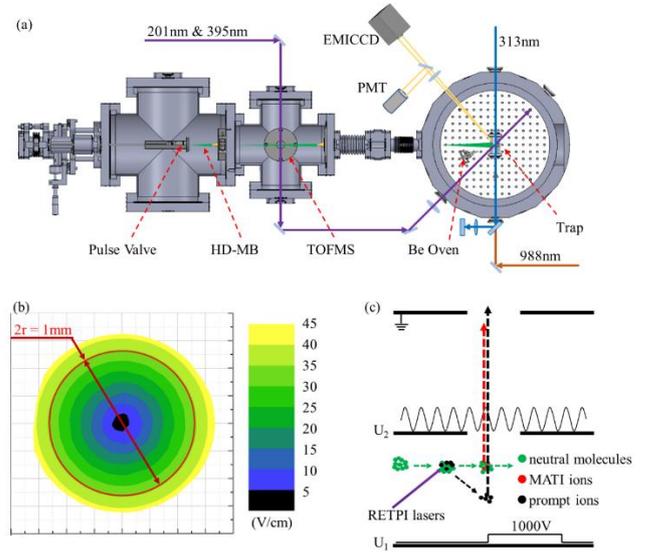

**FIG. 2.** (a) Schematic of the experimental setup. HD-MB, HD molecular beam; TOFMS, time-of-flight mass spectrometer. (b) Simulation result of the electric field strength distribution in the radial direction within 0.5 mm from the center of the ion trap. (c) Ionization of HD molecules in the TOFMS. $U_1$, pulse high voltage applied to the lower plate; $U_2$, rf voltage applied to the middle plate. Prompt ions are created by direct photoionization or electric field ionization caused by $U_2$. MATI ions are created by electric field ionization caused by $U_1$. The solid arrow represents the RETPI lasers, and dash arrows indicate the flight paths of molecules.

During direct photoionization of HD molecules in the ion trap, the state-unselected HD$^+$ ions can also be created by electric field ionization of neutral HD molecules excited into Rydberg states [Fig. 1(b)]. The ionization energy $E_{ion}$ is lowered by the electric trapping field of $A\sqrt{F/(\text{V cm}^{-1})}$, where $A$ represents the ionization capability of the electric field and $F$ is the electric field strength in units of V cm$^{-1}$ [47-48]. These effects are studied first by the TOFMS [Fig. 2(c)]. The Wiley-McLaren type TOFMS is composed of three electrode plates and a Microchannel Plate (MCP) [42]. The gap $d$ between the adjacent electrode plates is 2.4 cm and the flight distance of HD$^+$ ions between the upper electrode plate and the MCP is 57.6 cm. To create an electric field similar to that of the ion trap, a rf voltage $U_2 = U_{dc} + U_{rf}\cos(\Omega t)$ is applied to the middle plate, where $U_{dc}$ and $U_{rf}$ are the amplitudes of the dc and rf potentials respectively, and $\Omega = 2\pi \times 15.1$ MHz is the frequency of the rf field in the ion trap. The phase of rf electric field can be locked with the pulse RETPI lasers by a pulse generator (Quantum 9520). The pulse high voltage $U_1$=1000 V is applied to the lower plate to extract HD$^+$ ions to the MCP. The neutral HD populated in the lower (higher) Rydberg states are ionized by the electric field caused by $U_1$ ($U_2$). The ions ionized by $U_1$ and $U_2$ are the so-called mass-analyzed threshold ionization (MATI) ions and prompt ions, respectively. In addition, direct photoionized ions are indistinguishable from prompt ions in the flight time. Both prompt ions and MATI ions are observed in the TOFMS.

## III. RESULTS AND DISCUSSION

### A. Threshold photoionization of HD

The electric field ionization of the HD molecular beam is performed in the TOFMS. The dc electric field ionization is demonstrated by applying $U_2 = U_{dc} = 5$ V to the middle plate of the TOFMS. The MATI and prompt ion spectra of HD are shown in Fig. 3 [49]. The ionization energies observed from the MATI and prompt ion spectra are lower than the field-free ionization threshold $E_{ion}^0$ due to the applied electric field of $U_1$ and $U_2$, respectively. In the MATI ion spectrum, ion signal peaks in the range 25190 -25260 cm$^{-1}$ correspond to np0 Rydberg states converging to the $J^+ = 0$ rotational state of HD$^+$ [50]. It can be inferred that HD molecules populated in the intermediate state are excited to Rydberg states and then ionized by the electric field. The missing series of the np0 Rydberg states in MATI ion spectrum is due to the existence of np2 Rydberg states converging to the $J^+ = 2$ rotational state of HD$^+$. The neutral HD molecules excited to the np2 Rydberg states cannot be field-ionized due to the large frequency gap to the $X^+\ ^2\Sigma_g^+, v^+ = 0, J^+ = 2$ rotational excited state of HD$^+$. In the prompt ion spectrum, the np2 series Rydberg states are also observed in the range 25260 - 25360 cm$^{-1}$. Although neutral HD molecules can be photoionized directly by the ionization laser, some HD molecules are excited to np2 Rydberg states (above the ionization threshold) with longer neutral-state lifetimes, thereby reducing the prompt ion signal. No transition to np1 Rydberg states converging to the $J^+ = 1$ rotational state of HD$^+$ is observed in both MATI and prompt ion spectra, which gives no evidence of significant g/u symmetry breaking [51] or the g/u symmetry mixture of Rydberg states under the weak stray field in our experiments [52].

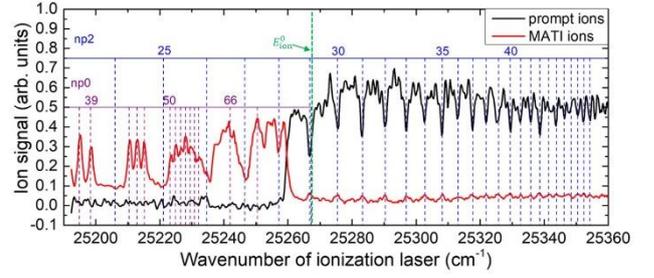

FIG. 3. Prompt and MATI ion spectra in the dc field. Dashed lines indicate np0 (purple) and np2 (blue) Rydberg states of neutral HD molecules; $IE_0$ is the field-free ionization threshold of neutral HD molecules accessible from the intermediate state.

The ionization energy $E_{ion}$ of neutral HD molecules in the intermediate state, derived from the onset of the prompt ion spectrum, is recorded as a function of the static electric field strength $F_{dc}$, where $F_{dc}$ is calculated by the amplitude $U_{dc}$ of the applied dc field and the gap $d$ between the adjacent plates of the TOFMS [Fig. 4(a)]. The $F_{dc}$ is scanned over the range 0.8 - 4.2 V/cm. The field-free ionization threshold of neutral HD molecules accessible from the intermediate state, $E_{ion}^0$ = 25267.1(2) cm$^{-1}$, and the ionization capability of the dc electric field, $A_{dc}$ = 3.6(2) cm$^{-1}$, are obtained by fitting the equation [47]

$$E_{ion} = E_{ion}^0 - A\sqrt{F/(\text{V cm}^{-1})}. \qquad (1)$$

The ionization energy $E_{ion}$ is further lowered by applying the rf electric voltage $U_2 = U_{dc} + U_{rf}\cos(\Omega t)$ to the middle plate of the TOFMS. The rf electric field strength $F_{rf}$, calculated by $(U_{dc} + U_{rf})/d$, is scanned over the range 21 - 31 V/cm, where $U_{dc}$ is kept constant at 5 V，and the amplitude of the rf electric field $U_{rf}$ is scanned over the range 45 - 70 V. Figure 4(b) shows the dependence of the ionization energy $E_{ion}$ on the rf electric field strength $F_{rf}$. Fitting with Eq. (1), the derived field-free ionization threshold of neutral HD molecules accessible from the intermediated state $E_{ion}^0$ = 25266.3(1.4) cm$^{-1}$ is consistent with the value in Ref. [50]. The derived $A_{rf} = 5.4(1)$ cm$^{-1}$, which represents the field ionization capability of the rf electric field, is in line with the typical value of 3.1 - 6.1 cm$^{-1}$ for the diabatic and adiabatic field-ionization [47]. Compared with the results of dc electric field, the larger error bars of the ionization energies derived from four measurements for each point are caused by the unlocked rf phase when the RETPI lasers intersect with the HD molecular beam.

To study the influence of the rf electric field phase on the ionization process, the prompt ion spectra are recorded when the ionization lasers are locked at different phases of the rf electric field [Fig. 5]. The "0°", "-30°", "-60°" and "-90°" denote the RETPI lasers intersecting with the HD molecules when the instantaneous electric field is 1, $\sqrt{3}/2$, 1/2 and 0 of the rf amplitude, respectively. In Fig. 5(a), in the range where the ionization laser frequency is below 25245 cm$^{-1}$, the ionization efficiency, indicated by the prompt ion signal intensity, is the highest when the phase is locked to "0°". The ionization efficiency is decreased as the phase is locked at "-30°", "-60°" and "-90°". This is due to the fact that HD molecules populated in Rydberg states are not field-ionized by the instantaneous rf electric field and deexcited before the rf electric field reaches its maximum. However, when the ionization photon energy is above 25245 cm$^{-1}$, the ionization efficiencies are similar for the phase locked at "0°", "-30°", "-60°". That is because the instantaneous rf electric field at these phases is sufficiently high to field ionize the Rydberg HD molecules. Moreover, when the frequency of the ionization laser is above the free-field ionization threshold $E_{ion}^0$, the ionization efficiency is almost not affected by the rf electric field phase [Fig. 5(b)]. In such a situation, HD$^+$ ions are mainly produced by direct photoionization.

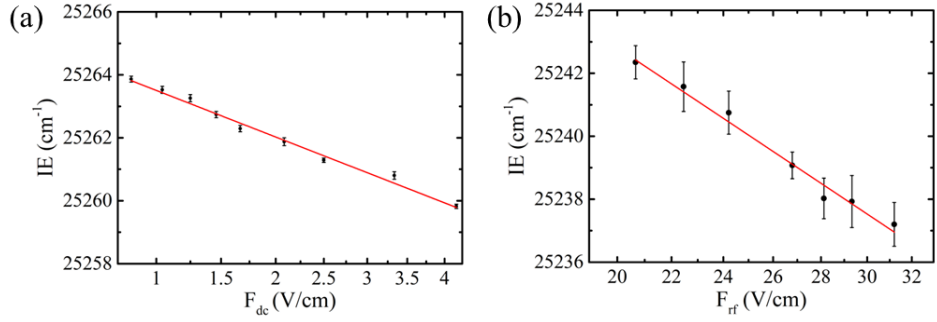

FIG. 4. The ionization energy $IE$ of neutral HD molecules in the intermediate state is recorded as a function of the electric field strength $F_{dc}$ (a) and $F_{rf}$ (b). All data points are the average of 4 measurement results. Vertical error bars represent the standard deviation. The red line is a fit according to the Eq. (1).

In general, the rotational ground state HD$^+$ ions can be generated by direct photoionization and the ionization efficiency is not affected by electric field when the photon energy of the ionization laser is above the field-free ionization threshold $E_{ion}^0$. However, to avoid neutral HD molecules being field-ionized to the $J^+ = 1$ rotational state due to the possible g/u symmetry mixing [52], the frequency of the ionization laser is set below the ionization energy of the $J^+ = 1$ rotational state (43.9 cm$^{-1}$ above the $E_{ion}^0$ [56]). Considering the influence of the electric field of the ion trap, the ionization energy of the $J^+ = 1$ rotational state is lowered by at most 31.9(4) cm$^{-1}$, which is calculated by the measured ionization capability of rf electric field $A_{rf}$ and the simulated maximal electric field strength in the ionization region of the ion trap. Therefore, the frequency of the threshold ionization laser is set between 25267.1(2) and 25279.1(4) cm$^{-1}$ for the generation of the rovibrational ground state HD$^+$ ions in our experiment.

### B. Detection of population of HD$^+$ ions

After the study performed in the TOFMS, rotational ground state HD$^+$ ions are generated in the ion trap using the [2+1'] RETPI. The experimental sequence for determining the rotational ground state population of HD$^+$ ions is depicted in Fig. 6(a). Typically, 21(1) HD$^+$ ions are created for the cumulation time of $\Delta t_c = 5$ s. HD$^+$ ions are trapped and sympathetically cooled by co-trapped laser-cooled Be$^+$ ions to form the HD$^+$/Be$^+$ bi-component Coulomb crystal [42,53]. After the cumulation, the frequency of the alternating current (ac) field used for secular motion resonance excitation of HD$^+$ ions was swept from 650 to

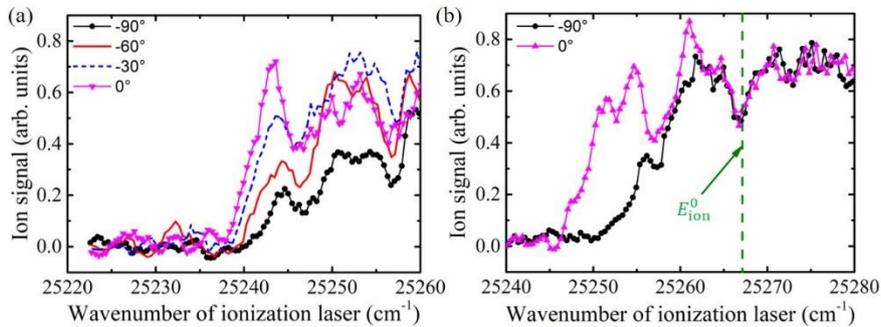

FIG. 5. The ionization efficiency of neutral HD molecules when the ionization lasers are locked at different phases of rf electric field. The vertical axis is the relative intensity of the ion signal recorded by MCP. The horizontal axis is the scanned frequency of the ionization laser. (a) The frequency of the ionization laser is below the field-free ionization threshold $IE_0$, and the amplitudes of electric fields $U_{dc}$ and $U_{rf}$ are set to 5 V and 65 V, respectively; (b) The frequency of the ionization laser is swept around the field-free ionization threshold $IE_0$, and the amplitudes of electric fields $U_{dc}$ and $U_{rf}$ are set to 5 V and 40 V, respectively.

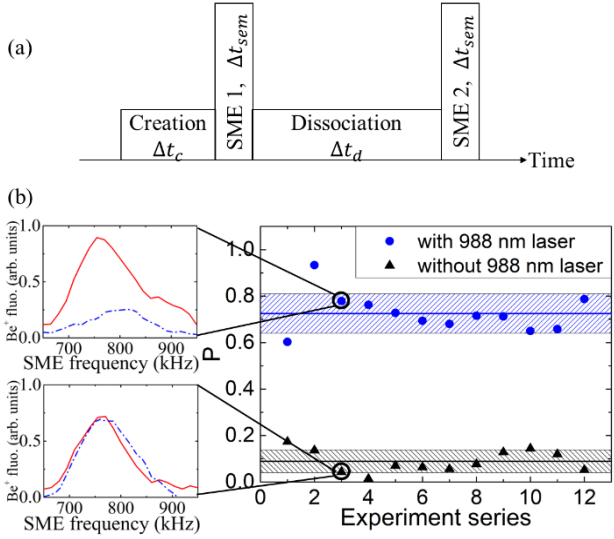

FIG. 6. Detection of the population of HD$^+$ ions by RETPI. (a) Experimental sequence for determining the population of HD$^+$ ions. The number of HD$^+$ ions is measured nondestructively before and after REMPD by secular motion excitation. SME, secular motion excitation. (b) The amount of HD$^+$ ions reduced during REMPD. The blue (black) data is obtained with (without) the 988 nm laser. The solid lines are the average of the results of 12 independent measurements and the shadows are the standard deviations. Inset: 313 nm fluorescence traces recorded using the PMT while exciting the secular motion of the HD$^+$ ions. The red (solid) and blue (dash dot) curves are taken before and after the [1+1'] REMPD.

950 kHz for the secular motion excitation time of $\Delta t_{sme} = 2$ s [42]. Then state-selected [1+1'] REMPD with the duration of $\Delta t_d$ using two continuous wave lasers at 988 nm and 313 nm was followed. Finally, the secular motion resonance excitation of HD$^+$ ions was repeated. Since the area under the 313 nm fluorescence trace $S$ at the secular motion resonance excitation of HD$^+$ ions is proportional to the number of HD$^+$ ions [54], the detection signal representing the reduced population of HD$^+$ ions is determined by $P = 1 - S_2/S_1$, where $S_1$ and $S_2$ represent the signals of the ion numbers in the ion trap before and after the REMPD, respectively [Fig. 6(b)]. The signal of HD$^+$ ion reduction caused by the collision of HD$^+$ with neutral HD and H$_2$ molecules in the background gas $P_{bg}$ is also detected in the same way but without using the 988 nm laser during the REMPD process [Fig. 6(b)]. So, the normalized rotational ground state population of HD$^+$ ions is obtained by $n_D(\Delta t_d) = (P - P_{bg})/(1 - P_{bg})$. For the dissociation duration of 10 s, $n_D(\Delta t_d = 10s) = 0.69(9)$ is obtained from $P = 0.72(9)$ and $P_{bg} = 0.09(5)$, which are both determined from 12 individual measurements [Fig. 6(b)].

## C. Evolution of rotational state populations

The time evolution of rotational state populations due to BBR is modeled by a set of rate equations (see Eqs. (2) and (3)). The rate equations are based on the Einstein A and B coefficients, which are derived from theoretical values of transition energies [55] and transition dipole moments [56] for HD$^+$ ions. For convenience, only the $J^+ < 6$ rotational states are included as there is almost no population in the $J^+ \geq 6$ rotational states of HD$^+$ ions at room temperature.

$$\frac{dn_i}{dt} = \sum_{j \neq i} R_{ij} \quad (2)$$

$$R_{ij} = \begin{cases} -B_{ij}\rho(v_{ij})n_i + B_{ji}\rho(v_{ij})n_j + A_{ij}n_j, & j > i \\ B_{ji}\rho(v_{ij})n_j - B_{ij}\rho(v_{ij})n_i - A_{ij}n_i, & j < i \end{cases} \quad (3)$$

where $n_i$ is the population of the $J^+ = i$ rotational state, $A_{ij}$, $B_{ij}$ and $B_{ji}$ are Einstein's coefficients of spontaneous emission, stimulated emission and absorption, and $\rho(v_{ij})$ is the spectral density of BBR. By solving Eq. (2) numerically, the ground-state population of HD$^+$ ions decays exponentially with a time constant $\tau \approx 11.5$ s as the system evolves back to the thermal equilibrium with room temperature. The rotational ground state population $n_m(t)$ of HD$^+$ ions generated after the m-th laser pulse evolves independently with time $t$ due to the BBR,

$$n_m(t) = n' + (n'' - n') \exp\left[-\frac{t - m/f_{rep}}{\tau}\right], \quad t \geq m/f_{rep}, \quad (4)$$

where $n''$ is the initial rotational ground state population of HD$^+$ ions, and $n' = 0.103(1)$ is the final rotational ground state population in the thermal equilibrium distribution of the ambient temperature and $f_{rep} = 10$ Hz is the repetition rate of the REPTI lasers. The population $n_c(t)$ of HD$^+$ ions created in a Coulomb crystal with the cumulation time of $\Delta t_c$ is the averaged population of all HD$^+$ ions generated by the number of $\Delta t_c \times f_{rep}$ laser pulses,

$$n_c(t) = [\sum_m n_m(t)]/(\Delta t_c \times f_{rep}). \quad (5)$$

Here, m ranges from 1 to $\Delta t_c \times f_{rep}$. The production rate of HD$^+$ ions in experiment is about 4.3 /s [42]. As it is difficult to determine which laser pulse leads to the production of each photoionized ion, we take the averaged effect and assume all laser pulses contribute equally to the production of the HD$^+$ ions.

According to the rate equations including the REMPD process of HD$^+$ ions, the time evolutions of the rotational ground and excited state populations of HD$^+$ ions $n_0$ and $n_e$ over time with $\Delta t_c = 5$ s and $\Delta t_d = 20$ s are shown in Fig. 7. The time step of $1/f_{rep}$ is used in the calculation. The rotational ground state population of HD$^+$ ions, $n_0$, is further reduced by the REMPD process (see point C - F in Fig. 7). Meanwhile, HD$^+$ ions at the rotationally excited states are deexcited to the ground state due to the BBR coupling and dissociated by the REMPD lasers. Their population, $n_e$, is reduced by $\Delta n_e$, which contributes to the rotational ground state population of HD$^+$ ions we detected. The rotational ground state population of HD$^+$ ions generated by RETPI, $n''$ and $n_c(\Delta t_c)$, can be deduced by combining the population evolution of HD$^+$ ions with the normalized rotational ground state population of HD$^+$ ions $n_D(\Delta t_d = 10 \text{ s})$ we detected. At the beginning of dissociation, the rotational ground state population is $n_c(t = \Delta t_c + \Delta t_{sem}) = n_D(\Delta t_d = 10 \text{ s}) - \Delta n_e(\Delta t_d = 10 \text{ s}) = 0.66(9)$ (see point C in Fig. 7). According to Eqs. (4) and (5), the initial ground state population $n''$ can be deduced as $0.93(12)$ (see point A in Fig. 7). The state-unselected HD$^+$ ions are mainly generated by the [2+1] resonance-enhanced multiphoton ionization (REMPI), which is agreed with the measured ratio of the yielding of HD$^+$ ions generated by the state-unselected [2+1] and state-selected [2+1']

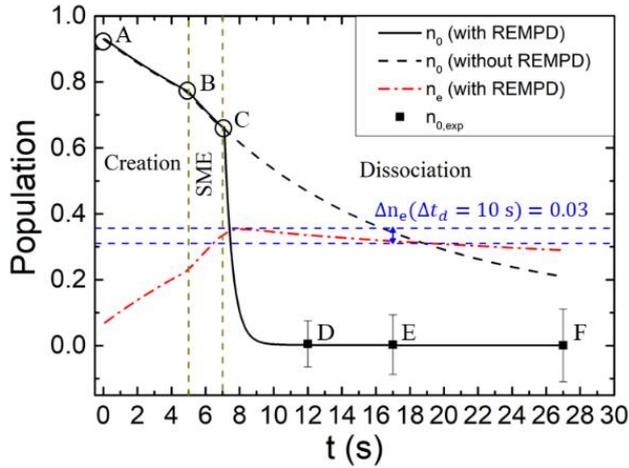

FIG. 7. Time evolution of the population of HD$^+$ ions. SME, secular motion excitation. $n_{0,exp}$ is the measured ground state population of HD$^+$ ions in the ion trap after dissociation, which is calculated by the equation $n_{0,exp}=1-n_D(\Delta t_d) - n_e$. The points D, E, F represent $n_{0,exp}$ at the dissociation time $\Delta t_d = 5$ s, 10 s, 20 s. The points D and F are both determined from 3 individual measurements, and the point E is determined from 12 individual measurements.

REMPI in the TOFMS [42]. The ground state population $n_c(\Delta t_c)$ of the HD$^+$ ionic Coulomb crystal we prepared for the cumulation time of $\Delta t_c = 5$ s is 0.77(8) (see point B in Fig. 7) and nearly a factor of 7.7 larger than the thermal equilibrium population at room temperature. In addition, the measured ground state populations $n_{0,exp}$, for $\Delta t_d = 5$ and 20 s, are consistent with the time evolutions of $n_0$ (with REMPD) (see points D and F in Fig. 7).

## IV. CONCLUSION

The rotational ground state HD$^+$ ions are produced in an ion trap by the [2+1'] RETPI, which has demonstrated the generation of state-selected polar molecular ions. The frequency of the threshold ionization laser is determined by taking into account the effect of electric field of the ion trap on the RETPI process, that avoids the production of state-unselected HD$^+$ ions via the electric field ionization. The rotational ground state population of HD$^+$ ions is detected by the state-selected [1+1'] REMPD. Combined with the evolution of the rotational state populations caused by BBR coupling, it is inferred that the initial ground state population $n''$ of HD$^+$ ions produced by RETPI is 0.93(12). The average ground state population $n_c(\Delta t_c)$ of HD$^+$ ionic Coulomb crystal prepared during the cumulation time of $\Delta t_c = 5$ s is 0.77(8). In fact, the time of the generation of the first ion is not precisely determined due to the fluctuation of the ionization laser energy. If the first ion is produced later than the starting point we assumed, the ground state population of HD$^+$ ions n" is overestimated. The average ground state population $n_c(\Delta t_c)$ can be further improved by increasing the ionization efficiency to shorten the cumulation time of the HD$^+$ ionic Coulomb crystal. The ionization efficiency can be enhanced by increasing the density of HD molecular beam and the laser energy while ensuring HD$^+$ ions to be generated by [2+1'] RETPI. In addition, a cryogenic Paul trap can be used to reduce the effect of BBR and increase the lifetime of the ground state population of HD$^+$ ions in a Coulomb crystal [57].

Our method can be further developed for the preparation of hyperfine state-selected HD$^+$ ions. The HD$^+$ ions with one unpaired electron of spin $s_e =1/2$, and the nuclei with spins $I_p = 1/2$ and $I_d = 1$ have four hyperfine states at zero magnetic field if the rotational angular momentum is $J^+ = 0$. Using an ionizing laser with a bandwidth of several hundred MHz, the HD$^+$ ions populated in the hyperfine ground state can be produced.

Our method is of interest in several directions. The high population $n''$ in the initially prepared ground state can facilitate to the quantum logic spectroscopy (QLS) experiments, which require fast loading of single molecular ions populated in the ground state [58]. We expect that the method presented here is particularly useful in the preparation of state-selected, translationally cold molecular ions for the study of cold ion-molecule collisions [59-60] and precision molecular spectroscopy [8-9]. In the near future, our state-selected HD$^+$ ions prepared by [2+1'] RETPI will be used for the recording of the rovibrational spectra of the transition $(v^+ = 0, J^+ = 0 \rightarrow v^+ = 6, J^+ = 1)$. The measured transition frequency can help to determine the fundamental constant of the proton-electron mass ratio (m$_p$/m$_e$) with the aid of the quantum electrodynamics (QED) theory.

## ACKNOWLEDGMENTS

We are grateful to Z.-C. Yan for careful and thorough reading of this manuscript. This work was supported by the National Key R&D Program of China (Grant No. 2021YFA1402103), the Strategic Priority Research Program of the Chinese Academy of Sciences (CAS) (Grant No. XDB21020200)) and the National Natural Science Foundation of China (Grant No. 91636216).

* Corresponding author: hesg@wipm.ac.cn
** Corresponding author: tongxin@wipm.ac.cn